\documentclass{mem}
\usepackage{natbib}\usepackage{txfonts}\usepackage{balance}
\usepackage{graphicx}
\usepackage[a4paper,breaklinks,dvipdfm]{hyperref}
\idline{0}{0}
\begin{document}
\def\teff{$T\rm_{eff }$}
\def\kms{$\mathrm {km s}^{-1}$}

\title{
The Formation History of the Ultra-Faint Dwarf Galaxies
}

   \subtitle{}

\author{
T.M. Brown\inst{1},
J. Tumlinson\inst{1},
M. Geha\inst{2},
E. Kirby\inst{3},
D.A. VandenBerg\inst{4},
J.S. Kalirai\inst{1},
J.D. Simon\inst{5},
R.J. Avila\inst{1},
R.R. Munoz\inst{6},
P. Guhathakurta\inst{7},
A. Renzini\inst{8},
H.C. Ferguson\inst{1},
L.C. Vargas\inst{2}, \&
M. Gennaro\inst{1}
}


\institute{
Space Telescope Science Institute, 3700 San Martin Drive,
Baltimore, MD 21218, USA
\email{tbrown@stsci.edu}
\and 
Astronomy Department, Yale University, New Haven, CT 06520, USA
\and
Department of Physics \& Astronomy, University of California Irvine,
4129 Frederick Reines Hall, Irvine, CA 92697, USA
\and
Department of Physics and Astronomy, 
University of Victoria, P.O. Box 3055, Victoria, BC, V8W 3P6, Canada
\and
Observatories of the Carnegie Institution of Washington, 
813 Santa Barbara Street, Pasadena, CA 91101, USA
\and
Departamento de Astronom\'ia, Universidad de Chile, 
Casilla 36-D, Santiago, Chile
\and
UCO/Lick Observatory and Department of Astronomy and 
Astrophysics, University of California, Santa Cruz, CA 95064, USA
\and
Osservatorio Astronomico, Vicolo Dell'Osservatorio 5, 
I-35122 Padova, Italy
}

\authorrunning{Brown et al.}

\titlerunning{Formation History of the UFDs}

\abstract{
We present early results from a {\it Hubble Space Telescope} survey of the
ultra-faint dwarf galaxies. These Milky Way satellites were
discovered in the Sloan Digital Sky Survey, and appear to be an
extension of the classical dwarf spheroidals to low luminosities,
offering a new front in the efforts to understand the missing
satellite problem.  Because they are the least luminous, most dark
matter dominated, and least chemically evolved galaxies known, the
ultra-faint dwarfs are the best candidate fossils from the early
universe.  The primary goal of the survey is to measure the
star-formation histories of these galaxies and discern any
synchronization due to the reionization of the universe.  We find that
the six galaxies of our survey have very similar star-formation histories,
and that each is dominated by stars older than 12~Gyr.
}
\maketitle{}

\section{Introduction}

One of the most prominent issues with the Lambda Cold Dark Matter
paradigm is the ``missing satellite problem'' -- the fact that the
predicted number of dark-matter halos exceeds the number of observed
dwarf galaxies \citep[e.g.,][]{moore99}.  As one possible solution,
\citet{bullock01} hypothesized that reionization suppressed star
formation in most dwarf spheroidals, making them unobservable in visible light.
Along these lines, \citet{ricotti05} proposed three different evolutionary
paths for dwarf galaxies: ``true fossils'' that formed most ($>$70\%)
of their stars prior to reionization, ``polluted fossils'' with
significant post-reionization star formation due to subsequent mass
accretion and tidal shocks, and ``survivors'' where star formation
largely began after reionization.  Around the same time, the discovery
of additional faint satellites and tidal debris around the Milky Way
\citep[e.g.,][]{willman05,zucker06,belokurov07} and Andromeda
\citep[e.g.,][]{zucker07,mcconnachie09} demonstrated that the census
of nearby dwarf galaxies was incomplete, and provided potential
examples of fossil galaxies.  Indeed, the first color-magnitude
diagrams (CMDs) of the ultra-faint dwarf (UFD) galaxies revealed that
they are dominated by old ($>$10~Gyr) stars
\citep[e.g.,][]{sand09,sand10,okamoto08,okamoto12,aden10,munoz10}.
Furthermore, the kinematics of these galaxies imply they are dominated
by dark matter \citep[e.g.,][]{kleyna05,munoz06,simon07}, and the
abundances indicate they are very metal-poor
\citep[e.g.,][]{frebel10,norris10,kirby11}.  Current simulations of
galaxy formation assume that most luminous dwarf galaxies formed their
stars over an extended period, but that most dark matter halos had
star formation truncated by reionization, or never formed stars at all
\citep[e.g.,][]{tumlinson10,munoz09,bovill09,bovill11a,bovill11b,koposov09}.

Against this backdrop, we proposed a large {\it Hubble Space Telescope
  (HST)} imaging program to better characterize the star formation
history (SFH) in a representative sample of six UFD galaxies.  Our
sample spans a range of galaxy luminosities, but avoids the brightest
UFDs (which may represent a transition to classical dwarf spheroidals) and the
faintest UFDs (which do not provide enough stars for a robust SFH
determination). The primary goal of the program is to measure relative
ages in these galaxies to better than 1~Gyr uncertainty. Such tight
age constraints are difficult when fitting optical CMDs for both age
and metallicity, but our program leverages the independent measurement
of the metallicity distribution function in each galaxy, as determined
from Keck/DEIMOS spectroscopy.  Relative ages among the UFD sample can
determine to what extent their SFHs have been synchronized
via reionization, and also tie their ages to those of other ancient
populations (e.g., Galactic globular clusters).  Because the {\it HST}
photometry extends well below the main-sequence turnoff (MSTO), these
data also provide insight into the initial mass function (IMF) in
these old, dynamically-unevolved populations.  We interpret our CMDs
using both empirical population templates observed in the same bands,
and also a high-fidelity isochrone library employing the latest
physics and extended to low metallicity ([Fe/H]~=$-4$).  Here, we
present a preliminary analysis of the {\it HST} data from each galaxy
in the program.

\section{Observations}

Our survey observed six UFD galaxies using two cameras on {\it HST}:
the Advanced Camera for Surveys (ACS) and the Wide Field Camera 3
(WFC3).  We used the F606W (broad $V$) and F814W ($I$) filters on each
camera, with ACS surveying an area centered on each galaxy
($202^{\prime\prime} \times 202^{\prime\prime}$ per tile), and WFC3
observing in parallel in the galaxy outskirts ($162^{\prime\prime}
\times 162^{\prime\prime}$ per tile).  For the nearby ($<$100~kpc)
satellites in the sample (Bootes~I, Coma Berenices, Ursa Major I), we
surveyed a relatively wide area (5--12 tiles in each camera) to a
shallow depth (2 or 3 orbits, faint limit of $V \sim 28$~mag). For the
distant ($>$130~kpc) satellites in the sample (Hercules, Leo~IV, and
Canes Venatici~II), we surveyed a narrow pencil beam (1 or 2 tiles in
each camera) more deeply (10--16 orbits, faint limit of $V \sim
29$~mag).  The varying depth and area of each survey was intended to
provide a sample of a few hundred stars in the vicinity of the MSTO 
with a photometric precision of $\sim$0.01~mag, thus providing tight
constraints on age when coupled with independent knowledge of the 
metallicity distribution from Keck.

\begin{figure*}[t!]
\resizebox{\hsize}{!}{\includegraphics[clip=true]{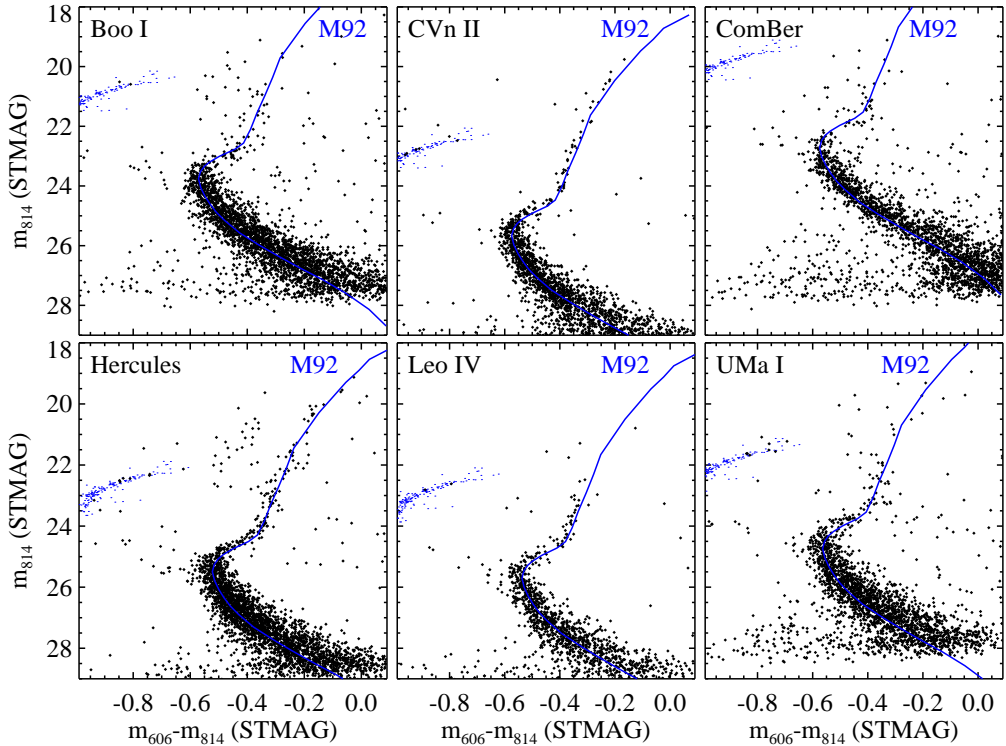}}
\caption{\footnotesize
The CMD of each UFD in our {\it HST} survey ({\it black points}). For
comparison, we show the ridge line ({\it blue curve}) and HB locus
({\it blue points}) from the CMD of M92 \citep{brown05}, an ancient 
\citep[12.75~Gyr;][]{vandenberg13} metal-poor \citep[$-2.3$;][]{harris96}
globular cluster observed in the same bands on the same {\it HST}
camera but shifted in color and magnitude to account for differences
in reddening and distance. The UFDs are clearly dominated by ancient
metal-poor stars.  The area and depth of observations for each galaxy
were varied to provide a few hundred stars in the vicinity of the MSTO, and
are largely responsible for the varying amount of field contamination
present in each CMD.  Despite this field contamination, a prominent BSS
can be discerned in each galaxy CMD.
}
\end{figure*}

The ACS images were processed through the standard pipeline, including
a pixel-based correction for charge transfer inefficiency
\citep[CTI;][]{anderson10}.  The images were then coadded using the
{\sc drizzle} package \citep{fruchter02}, including correction for
geometric distortion, resampling of the point spread function (PSF),
and masks for cosmic rays and detector artifacts.  We performed
PSF-fitting photometry using the DAOPHOT-II package \citep{stetson87},
producing a catalog for each galaxy in the STMAG system: $m= -2.5
\times $~log$_{10} f_\lambda -21.1$.  Taking a star at an old (13 Gyr)
metal-poor ([Fe/H]=$-2.5$) MSTO for reference, the offsets between the
STMAG system and the familiar Johnson system (tied to Vega) are $I =
m_{814} - 1.31$~mag and $V = m_{606} - 0.14$~mag.  The catalogs were
cleaned of background galaxies and stars with poor photometry.  We
performed artificial star tests to evaluate photometric scatter and
completeness, including CTI effects, with the same photometric
routines used to create the photometric catalogs.  The CMD for each
UFD in our survey is shown in Figure 1.

\section{Analysis}

Although the populations of the UFDs span more than 2~dex in
metallicity \citep[e.g.,][]{kirby11},
the bulk of the stars fall at
metallicities of $-2 >$~[Fe/H]~$> -4$.  The population of the ancient
\citep[12.75~Gyr;][]{vandenberg13}, metal-poor
\citep[$-2.3$;][]{harris96} globular cluster M92
is thus an important point of comparison.
In Figure 1, we show compared to each UFD CMD the ridge line and
horizontal branch (HB) from the CMD of M92, observed in the same ACS
bands \citep{brown05}.  It is clear that the CMD of each UFD
looks like the CMD of an ancient globular cluster,
and that each is dominated by an old (age~$>$12~Gyr)
population of metal-poor ([Fe/H]~$< -2$) stars.  A prominent blue straggler
sequence (BSS) is present in each galaxy -- ubiquitous in old
populations.

The luminosity difference between the MSTO and the HB is a well-known
age indicator -- the MSTO luminosity decreases with age, while the the
HB luminosity is relatively constant with age.  For populations with
similar chemical compositions (as is the case here), the color
difference between the MSTO and the base of the red giant branch (RGB)
is also an age indicator \citep{vandenberg90}, because the MSTO
becomes redder at increasing age while the RGB is relatively
insensitive to age.  To empirically demonstrate the extent of
synchronization in the SFHs of the UFDs, we show in Figure 2 the
composite CMD from our entire sample, with each UFD shifted to the
frame (distance and reddening) of Hercules.  The fact that each UFD
CMD is similar to that of Hercules indicates that their SFHs are
largely synchronized, with mean ages agreeing to $\sim$1~Gyr; better
constraints on the degree of synchronization will come from refinements to 
the quantitative SFH fitting, which we are pursuing.

Using synthetic CMD analysis, \citet{brown12} found that the mean age
of Hercules was only 0.1~Gyr younger than that of M92.  The absolute
age of M92 itself is driven by uncertainties in both distance and
[O/Fe]; indeed, the 12.75~Gyr age of \citet{vandenberg13} is nearly 1
Gyr younger than previous estimates, due primarily to the adoption a
higher [O/Fe] and a larger distance modulus.  Our synthetic CMD
analyses of each UFD CMD are underway, and indicate that the bulk of
the population is older than 12~Gyr in each galaxy.  However, we are
still exploring how best to handle the membership probabilities for
the spectroscopic sample.  Specifically, minority populations of
relatively metal-rich stars ($-1 >$~[Fe/H]~$> -2$) in the metallicity
distribution function may represent field contamination or a small
continuation of star formation beyond the dominant burst.  We are also
investigating the IMF of each galaxy; our analysis of the first two
galaxies observed in our survey (Hercules and Leo~IV) found that their
IMFs are flatter than those in the Galactic field and luminous dwarf
satellites \citep{geha13}.

\begin{figure}[t!]
\resizebox{\hsize}{!}{\includegraphics[clip=true]{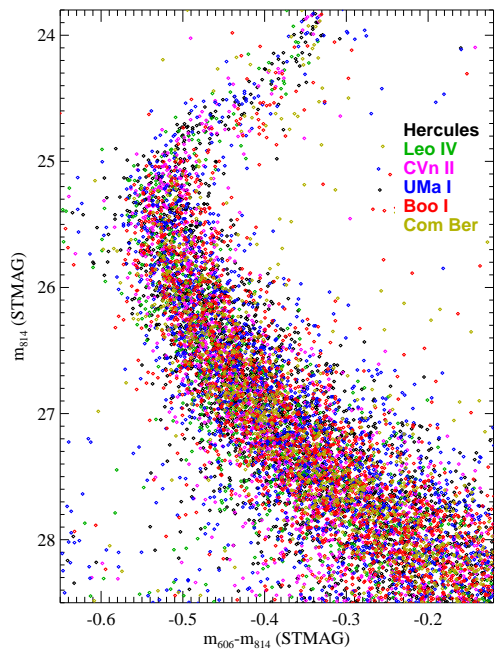}}
\caption{\footnotesize
The composite CMD of all six UFD galaxies in our survey ({\it
  points}), with the CMDs of 5 galaxies ({\it colored points}) shifted
to the frame of Hercules ({\it black points}).  The CMDs are nearly
indistinguishable from each other, with similar luminosity differences
between the MSTO and HB, and similar color differences between the
MSTO and RGB base.  These similarities imply that their star formation
histories are synchronized to $\sim$1~Gyr.
}
\end{figure}

\section{Summary}

The {\it HST} photometry of these six UFD galaxies demonstrates that
they are dominated by truly ancient stars ($>$12~Gyr old), and that
their SFHs are largely synchronized (with our initial estimates
indicating agreement at the level of $\sim$1~Gyr).  The uniformly
ancient populations of the UFDs stand in contrast to other dwarf
galaxies in the local universe, most of which formed the bulk of their
stars prior to $z \sim 1$ \citep{weisz11}.  The synchronization of SFH
in the UFDs suggests that a global phenomenon is at work, such as
reionization, instead of other stochastic mechanisms for truncating
the SFH, such as gas depletion or supernova feedback.  The distinction
between these interpretations will be clarified as we refine the SFH
fits.  We are improving several aspects of our analysis, including the
assumed variation in [O/Fe] as a function of [Fe/H], and the selection
criteria used to construct the metallicity distribution function for
each galaxy.  These improvements will determine to what extent any
residual star formation may have continued beyond the dominant burst
in each galaxy.  Given what we know now, the UFDs are the best
examples of fossil galaxies, and lend credence to the hypothesis that
many of the ``missing satellites'' remain undiscovered because they
have formed little to no stars.
 
\begin{acknowledgements}
Support for GO-12549 was provided by NASA through a grant from
STScI, which is operated by AURA, Inc., under NASA contract NAS
5-26555.  A.R. acknowledges support from ASI via grant
I/009/10/0.  R.R.M. acknowledges support from the GEMINI-CONICYT Fund,
allocated to the project N$^{\circ}32080010$.
\end{acknowledgements}

\bibliographystyle{aa}

\end{document}